%% file: main.tex
\documentclass[runningheads]{llncs}
\input{command}

\begin{document}
	
\input{contribution}

\begin{abstract}
	Heterogeneous clusters with nodes containing one or more accelerators, such as GPUs, have become common. While MPI provides inter-address space communication, and OpenCL provides a process with access to heterogeneous computational resources, programmers are forced to write hybrid programs that manage the interaction of both of these systems. This paper describes an array programming interface that provides users with automatic and manual distributions of data and work. Using work distribution and kernel {\em def} and {\em use} information, communication among processes and devices in a process is performed automatically. By providing a unified programming model to the user, program development is simplified.
	
\end{abstract}

\section{Introduction}
\label{s:intro}
\input{intro}

\section{Design of the \name\ Interface}
\label{s:design}
\input{design}

\section{Implementation}
\label{s:impl}
 \input{implemention}

\section{Experimental Results}
\label{s:eval}
\input{eval}

\section{Related Work}
\label{s:related}
\input{related}
\section{Conclusions and Future Work}
\label{s:conclusion}
\input{conclusion}

\section*{Acknowledgments}
This material is based upon work supported by the National Science Foundation under Grant No. CNS-1405954, and used the Extreme Science and Engineering Discovery Environment (XSEDE), which is supported by National Science Foundation grant number ACI-1548562. Any opinions, findings, and conclusions or recommendations expressed in this material are those of the author(s) and do not necessarily reflect the views of the National Science Foundation or XSEDE.  We also thank Prof. Jeffrey M. Siskind and Purdue ITaP for the use of their resources.

\bibliographystyle{splncs04}
\bibliography{bib/paper}

\end{document}

%% file: command.tex
\newcommand{\name}{HDArray}

\newcommand{\gdefsend}{{sGDEF}}
\newcommand{\gdefrecv}{{rGDEF}}

\newcommand{\sendmessage}{{SENDMSG}}
\newcommand{\receivemessage}{{RECVMSG}}
	
\newcommand{\funcinit}{{\tt \name Init}}
\newcommand{\funccreate}{{\tt \name Create}}
\newcommand{\funcapplykernel}{{\tt \name ApplyKernel}}
\newcommand{\funcexit}{{\tt \name Exit}}
\newcommand{\funcread}{{\tt \name Read}}
\newcommand{\funcwrite}{{\tt \name Write}}
\newcommand{\funcreduce}{{\tt \name Reduce}}
\newcommand{\funcshow}{{\tt \name ShowDeviceInfo}}
\newcommand{\funcpartition}{{\tt \name Partition}}
\newcommand{\partitiontype}{{\tt PART\_T}}

\newcommand{\funcabsuse}{{\tt \name SetAbsoluteUse}}
\newcommand{\funcabsdef}{{\tt \name SetAbsoluteDef}}
\newcommand{\functrapezoiduse}{{\tt \name SetTrapezoidUse}}
\newcommand{\functrapezoiddef}{{\tt \name SetTrapezoidDef}}

\newcommand{\funcproto}[1]{{\tt #1}}

\newcommand{\Samcomment}[1]{}
\newcommand{\Hyundokcomment}[1]{}
\newcommand{\Okwancomment}[1]{}

\usepackage{cite}

\usepackage[pdftex]{graphicx}
\graphicspath{{fig}}
\DeclareGraphicsExtensions{.pdf,.jpeg,.png}

\usepackage[cmex10]{amsmath}
\usepackage{algorithm}
\usepackage{algorithmicx}
\usepackage{algpseudocode}
\algrenewcommand\textproc{}

\usepackage{array}

\usepackage[caption=false,font=footnotesize]{subfig}

\usepackage{url}

\usepackage[english]{babel}
\usepackage{underscore}

\usepackage{tabularx}

\usepackage{listings}
\usepackage{color}
\usepackage[dvipsnames]{xcolor}

\newcommand*{\textsmallunderscore}{%
	\begingroup
	\fontencoding{T1}\fontfamily{lmtt}\selectfont
	\textunderscore
	\endgroup
}

\usepackage{microtype}
\usepackage[T1]{fontenc}
\usepackage[scaled]{beramono}

\usepackage{verbatim}

\usepackage{enumitem}

\usepackage[export]{adjustbox}

\makeatletter
\newcommand\notsotiny{\@setfontsize\notsotiny\@vipt\@viipt}
\makeatother

\makeatletter
\renewcommand\paragraph{\@startsection{paragraph}{4}{\z@}%
	{-8\p@ \@plus -4\p@ \@minus -4\p@}
	{-0.5em \@plus -0.22em \@minus -0.1em}%
	{\normalfont\normalsize\itshape}}
\makeatother

%% file: contribution.tex
%

\title{ \name: Parallel Array Interface for Distributed Heterogeneous Devices}
\titlerunning{\name: Parallel Array Interface for Distributed Heterogeneous Devices}

\author{Hyun Dok Cho\inst{1}\protect\footnote{This work was done while at Purdue University.} \and
	Okwan Kwon\inst{1} \and
	Samuel P. Midkiff\inst{2}}

\authorrunning{H. Cho et al.}
%

\institute{
	NVIDIA Corporation, Santa Clara, CA 95050, USA\\
	\email{\{hyundokc,okwank\}@nvidia.com}
	\and
	Purdue University, West Lafayette, IN 47907, USA\\
	\email{smidkiff@purdue.edu} 
}


\maketitle              

%% file: intro.tex
As GPU programming becomes more mainstream, both large and small scale multi-node systems  with one or more GPUs per node have become common. These nodes, however, complicate already messy distributed system programming by adding  MPI~\cite{gropp1999using} on top of proprietary host-GPU mechanisms. Developers must maintain two programming models: one for intra-process communication among devices and one for inter-process communication across address spaces.

%
Several systems have improved the programmability of multi-node systems with accelerators. 
%
%
SnuCL~\cite{kim2012snucl, kim2016distributed}, dCuda~\cite{gysi2016dcuda}, and IMPACC~\cite{kim2016impacc} support transparent access to accelerators on different nodes, and PARRAY~\cite{Chen:2012:PUA:2145816.2145838} and Vi{\~n}as \textit{et al}.~\cite{vinas2016towards} provide high-level language abstractions and flexible array representations. 
Programmers can develop high-performance applications but must manage low-level details of accelerator programming or provide explicit communication code. 
Partitioned Global Address Space (PGAS) platforms for accelerators, XMP-ACC~\cite{Lee:2011:EXP:2238356.2238410}, XACC~\cite{7081675}, and Potluri \textit{et al}.~\cite{potluri2013extending}, relieve programmers from dealing with data distribution, but data is strongly coupled to threads, making performance tuning more difficult. Finally, compiler-assisted runtime systems, Hydra~\cite{sakdhnagool2015hydra} and OMPD~\cite{ Kwon:2012:HAO:2145816.2145827}, propose a fully automatic approach that allows OpenMP programs to run on accelerator clusters, presenting an attractive alternative for developing repetitive, regular applications, but the distribution of work and data are limited by OpenMP semantics and expressiveness. 

In this paper, we describe the Heterogeneous Distributed Array (\name) interface and runtime system.  \name\ targets program execution on cluster-sized distributed systems with nodes containing one or more accelerators, i.e., devices.  Work is done as OpenCL work items, and \name\ provides ways to explicitly and implicitly partition work onto devices.  

\name\ also provides a way for the data used by the work on a device to be specified.  
The data read and written is typically relative to work items and can be specified either using offsets from the work item, or with an absolute specification of the data. 
\name\ then tracks the data defined and used by each work item, which allows communication to be generated automatically, since, in race-free programs, \name\ knows where the last written copy of a datum is, and who needs that value.  Importantly, data is not explicitly distributed and is not bound to, or owned by, a work item, but flows from its defining process and device to the process and device where it is needed.  
   
Finally, \name\ allows work to be repartitioned at any point in the program. This flexibility allows a programmer to optimize the work distribution and its necessary communication without any changes to the kernel code.
\if(0)
In this paper, we describe the Heterogeneous Distributed Array ({\em \name}) interface and runtime system that provides automatic communication and simple tuning for high performance.
\name\ targets program execution on cluster-sized distributed systems whose nodes contain one or more accelerators. OpenCL~\cite{5457293} is used to create work on {\em devices}, either accelerators or cores.  
\name\ APIs and annotations provide a global address space programming model to enable both automatic communication \Samcomment{is there a reason we don't say "automatic communication" instead of "high level abstractions of communication"?}\Hyundokcomment{Agree. We should say automatic. Updated.} and flexible distribution options.

The \name\ provides automatic communication when supplied with annotations providing work distribution and the data needed by the work. The annotations allow the specifications of the data read and written during kernel invocations. The specifications can also provide \emph{offset}s which refer to data boundary relative to each OpenCL work item, and for a large class of kernel code, these offsets are easy for programmers to provide.  As well, \name\ supports explicit and implicit partitioning of work item regions across devices. With offset and partition information, the \name\ runtime automatically generates efficient communication. This relieves programmers from handling explicit communication, allowing them to focus on kernel programming. Furthermore, \name~allows programmers to distribute data and work separately through APIs using a partitioned work item region ID. Therefore, there is no data ownership, and programmers can selectively use different partitions for both data and work at any program point. This flexibility allows programmers to optimize work distribution, and therefore communication, without any modification of existing kernel code. 
\fi

To summarize, our contributions are 
\begin{enumerate}
	\item A novel and easy programming model that can efficiently run on distributed heterogeneous devices, enabling flexible work distribution, with data flowing to the work that needs it.
	\item A fully automatic runtime communication generation scheme and its implementation.
	\item A flexible user interface that allows manual tuning of work distribution for high performance and enabling automatic communication. 
	\item Experimental results showing good performance and speedups on small clusters with eight nodes and 1 to 32 GPUs.
\end{enumerate}

The rest of the paper is organized as follows. 
Sections~\ref{s:design}, ~\ref{subs:ui}, and~\ref{s:impl} describe the design and implementation of the \name~interface. Section~\ref{s:eval} presents a performance evaluation of the \name~runtime system. Section~\ref{s:related} discusses related work, and conclusions and future work follow in Section~\ref{s:conclusion}.

%% file: design.tex
\begin{figure}[htb]
	\centering 
	\vspace{-15pt}
	\subfloat[ The \name\ frontend]{\label{sf:sysA}  \includegraphics[width=0.47\textwidth]{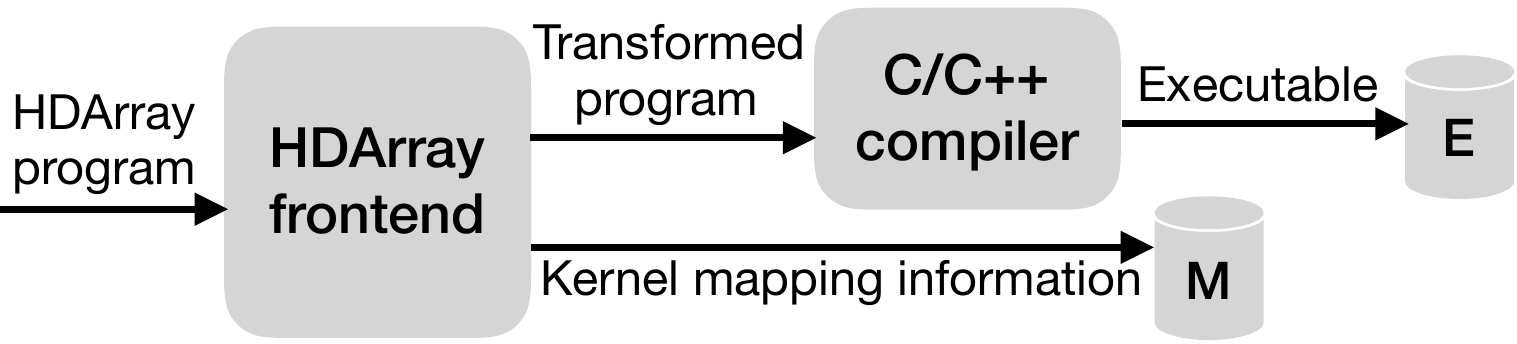}}%
	\vspace{-2pt}
	\hskip 0.1truein
	\subfloat[ The \name\ runtime]{\label{sf:sysB}\includegraphics[width=0.46\textwidth]{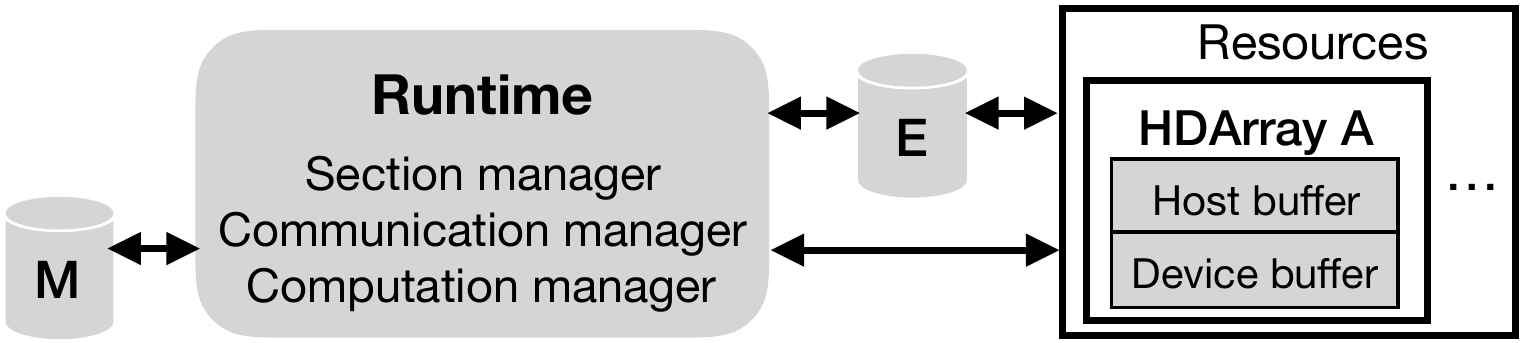}}
	\caption{An overview of the \name\ system within a single process.}
	\label{fig:overview}
	\vspace{-10pt}
\end{figure}
We now present the design of the \name\ interface and its runtime system as shown in Fig.~\ref{fig:overview}.
The central structure, and concept, of the \name\ system is the \name\ itself.  
The \name\ encapsulates a {\em host buffer} and {\em device buffer} by keeping necessary states for communication among processes and kernel computation.
%
The \name\ system provides a collection of APIs and annotations that the programmer uses to access the features of the \name\ system.  These APIs and annotations are translated by the \name\ frontend into calls, arguments, and initialization files (M) for use by the \name\ runtime.

\subsection{\name\ Structures}
\label{subs:structures}

Each MPI process that maps to a single OpenCL device maintains \name s and their structures. Each \name\ contains the necessary state for the system to automatically generate communication and manage data distributions.

%
%

Host and device buffers contain the distributed data for an \name\ for a host and device in the \name\ program, respectively. Host buffers reside in the process memory of the host, and device buffers reside on a device. 

An \name\ contains sets of array sections: {\em global}, i.e., across all kernels, definition sections ({\em GDEF}); {\em local}, i.e., for a particular kernel, definition sections ({\em LDEF}); and local use sections ({\em LUSE}). 
These sets are summarized by one or more sections of [LB:UB] that give the lower and upper bounds of the array sections for all processes. 
GDEF is a set of written sections not propagated to different processes, and 
two types of GDEFs are maintained: \gdefsend\ and \gdefrecv. 
\gdefsend\ for a process $ p $ describes \name\ elements that $ p $ has written, but not sent, to other processes $ q $,
i.e., the elements for which $ p $ describes the coherent copy that must be sent to other processes that use those values.
\gdefrecv\ for process $ p $ describes elements of the \name\ that $ p $ has not received from $ q $.
LUSE/LDEF is the set of sections each process reads/writes in the kernel.

\name\ programs are SPMD programs, and each process maintains coherent local copies of the aforementioned four sets for all processes,  
and thus each process knows the array access information of the other processes.
%
%
All LUSE, LDEF, and GDEF sets are empty when an \name\ is created. LUSE (LDEF) is updated by \name\ annotations and APIs, as discussed in Section~\ref{subs:ui}, and GDEF is updated as a function of itself, LUSE, and LDEF, which we describe in Section~\ref{subs:comm_gen}. 
%

As well, each process maintains a history of  local and global sections, and sections to communicate for each \name\ of a kernel. The runtime maintains the history to reduce the overhead of determining communication by avoiding an expensive data flow evaluation if possible, as described in Section~\ref{subs:optimization}.

\subsection{Communication Generation using GDEF, LDEF, and LUSE}
\label{subs:comm_gen}
The runtime communicates elements of \name s immediately before a kernel launch. Intersecting the GDEF and LUSE sets allows a process to determine which processes have elements of an \name\ that it will use in a kernel call, and therefore which elements it must communicate with that process.  
\begin{figure*}[htb]
	\centering
	\subfloat{\includegraphics[max width=4.6in]{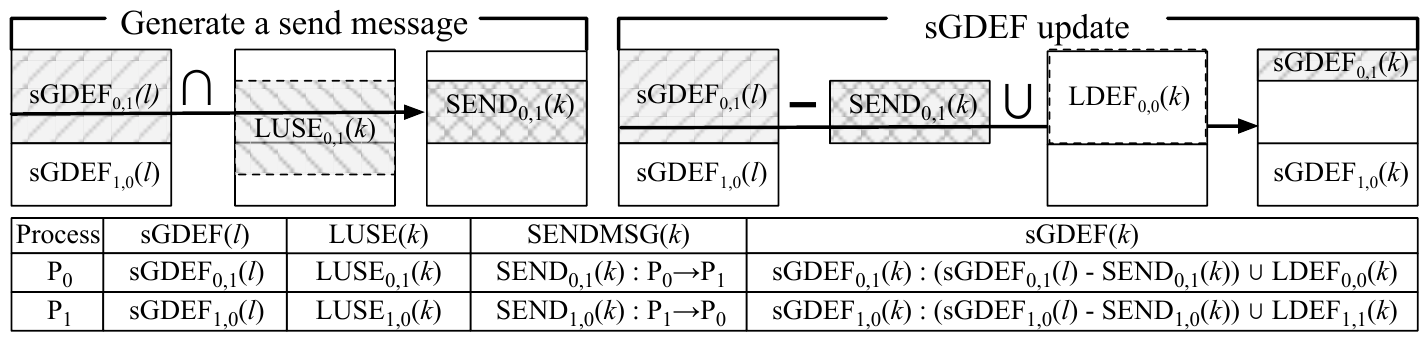}}%
	\vspace{-10pt}
	\caption{\gdefsend\ of \emph{use} \name\ for Process 0 before and after kernel call (\gdefsend$_{0,1}$($ l $) and \gdefsend$_{0,1}$($ k $) respectively). The system intersects \gdefsend\ with LUSE, sends the intersection from P\textsubscript{0} to P\textsubscript{1}, and updates \gdefsend. Note that LDEF$_{0,0}$($ k $) is NULL.}
	\label{fig:designcomm}
	\vspace{-13pt}
\end{figure*}
Fig.~\ref{fig:designcomm} shows the example of the intersection with two processes.
Let $ k $ be the index of a kernel call and $ l $ the preceding kernel call of $ k $.
For process $ p $, \gdefsend$_{p,q}$($ l $)  and \gdefrecv$_{p,q}$($ l $) are sets of sections that are live before the kernel call $ k $. 
LUSE$_{p,p}$($ k $) is local uses by process $ p $ for the kernel call $ k $, whereas LUSE$_{p,q}$($ k $) represents  local uses of other process $ q $ for $ 0 \leq q \leq nprocs - 1, q \neq p $. 
The communication messages of $ p $ to send and receive for the kernel call $ k $ are then generated as:
{\small
\begin{align}
{\it \sendmessage}_{p,q}(k) = {\it \gdefsend}_{p,q}(l)  \cap {\it LUSE}_{p,q}(k) \label{eq:intersectionsend}\\
{\it \receivemessage}_{p,q}(k) = {\it \gdefrecv}_{p,q}(l)  \cap {\it LUSE}_{p,p}(k) \label{eq:intersectionrecv}
\end{align}
}%
After communication and kernel execution, GDEF sets for each \name\ for kernel call $ k $ must be updated to avoid redundant communication for used \name s, and detect new communication in the future kernel call $ k + 1 $ for defined \name s. These updates can be calculated together as Eqns.~\ref{eq:update_gdef_send} and~\ref{eq:update_gdef_recv}. 
{\small
\begin{align}
{\gdefsend}_{p,q}(k) = ({\gdefsend}_{p,q}(l) - {\sendmessage}_{p,q}(k)) \cup {\it LDEF}_{p,p}(k) \label{eq:update_gdef_send}\\
{\gdefrecv}_{p,q}(k) = ({\gdefrecv}_{p,q}(l) - {\receivemessage}_{p,q}(k)) \cup {\it LDEF}_{p,q}(k) \label{eq:update_gdef_recv}
\end{align}
}%
Fig.~\ref{fig:designcomm} shows the example of the \gdefsend\ update. 
The \sendmessage\ and \receivemessage\ for the kernel call $ k $ are subtracted from \gdefsend\ and  \gdefrecv\ of kernel call $ l $, respectively. These removes communicated data from \gdefsend\ and \gdefrecv\ of the kernel call $ l $. The results of the subtractions are then unioned with LDEF sets to update \gdefsend\ and \gdefrecv\ for kernel call $ k $. 
Similar to LUSE, LDEF$_{p,p}$($ k $) and LDEF$_{p,q}$($ k $) sets represent local definitions by process $ p $ and $ q $ for kernel call $ k $, respectively. 


\section{\name\ Programming Interface}
\label{subs:ui}
The \name\ programming interface has two types of specifications. First, a single pragma of the form {\tt \#pragma hdarray [clauses]} allows user-defined hints for generating LUSE and LDEF to find data to be accessed, and partitioning work item regions to distribute work.  The functionality is contained in the clauses, described next.  Second, \name\ provides library functions, hiding low-level details of distributed device programming, described below.
Hereafter, we use the terms work item and thread interchangeably. 


%
\label{subs:clauses}
Table~\ref{tab:clauses} lists the available annotation clauses. Five clauses exist: {\em use}, {\em def}, {\em use@}, {\em def@},  and {\em partition}.  We now explain each of these.

\begin{table}[tb]
	\notsotiny
	\renewcommand{\arraystretch}{1.3}
	\caption{Core \MakeLowercase{\name} directive clauses}
	\begin{center}
		\vspace{-13pt}
		\resizebox{\columnwidth}{!}{%
			\begin{tabular}{|l|l|}
				\hline
				\bf Clause & \bf Description\\
				\hline
				\emph{use} (array name, offset) & Declare offset(s) of an array to be used.\\
				\hline
				\emph{def} (array name, offset) & Declare offset(s) of an array to be defined.\\
				\hline
				\emph{use@} (array name) & Declare absolute sections(s) of an array to be used.\\
				\hline
				\emph{def@} (array name) & Declare absolute sections(s) of an array to be defined.\\
				\hline
				\emph{partition} (partition ID, dimension, dev:ID, region) & Manually partition work item regions to devices.\\
				\hline
			\end{tabular}
		}
		\vspace{-10pt}
	\end{center}
	\label{tab:clauses}
\end{table}
%

\paragraph{Offset Clauses: \textbf{\textit{use}} and \textbf{\textit{def}}.} 
Each \name\ accessed in a kernel can be a \emph{use}, \emph{def}, or both. These clauses specify elements of arrays, as offsets relative to work items, that will be read and/or written by a single work item. The offsets can be used when a kernel's array access pattern is relative to a work item, which is the most common kernel programming pattern. If the offsets are specified, the system derives LDEF and LUSE from the offset and partitioned work item region for each process. 
An offset can be any integer, e.g., ``0'' indicates the current position of an array element relative to the work item index. It also describes a direction with $``+"$ or $``-"$; e.g., $(0, -1)$, referring to the previous elements of the same row. An $``*"$ denotes all elements of the array in the dimension of interest, e.g., $(0, *)$ denotes all elements in a row of a 2-dimensional array.

\paragraph{Absolute Section Interface Clauses: \textbf{\textit{use@}} and \textbf{\textit{def@}}.}
\label{subs:absclause} 
When it is difficult to represent offsets, e.g., a kernel's access pattern is not relative to a work item or non-rectangular, one can use the \emph{absolute section interface} with \emph{use@} and/or \emph{def@} clauses and APIs.
Unlike offsets, absolute sections are the coordinates of the [lowerbound, upperbound] of each dimension of an up to three dimensional item.  The absolute section interface clauses inform the system to bypass LUSE/LDEF updates (Fig.~\ref{fig:applykernel}).  Instead, users call absolute section interface APIs, e.g., \funcabsuse, described in Section~\ref{ssub:api}, to set LDEF and LUSE.  The APIs allow users to specify multiple absolute sections for each device, allowing non-rectangular regions or different access patterns for each device to be described, and enable fine tuning of communication. 

\paragraph{Partition Clause: \textbf{\textit{partition}}.} 
\name\ provides an  \funcpartition\ library call that allows a ROW, COL or BLOCK partition to be specified.  \name\ will automatically partition the work evenly across processes and devices in the specified manner.  A ROW partitioning is shown in Listing~\ref{code:host}.  The API then returns  a unique partition ID, which is used, repeatedly if necessary, to execute kernels with the partitioned work item. This partition ID, along with {\it use} and {\it def} information, allows the system to know data needed by the work done by the kernel on each device. 


\begin{lstlisting}[caption={Manually partitioned Correlation host code.},label={code:partition}, frame=lines, breaklines=true, xleftmargin=2em, framexleftmargin=1.5em, numbers=left, numberstyle=\tiny, numbersep=5pt, captionpos=b, float=tb, belowskip=-3.5em, aboveskip=-0em]
...
#pragma hdarray partition(part0,        (10240,10240),\
                          dev:0,   (0,3008),(0,10240),\  
                          dev:1,(3008,7232),(0,10240))
...
HDArrayApplyKernel("corr_ker1", part0, ... );
HDArrayApplyKernel("corr_ker2", part0, ... );
...
\end{lstlisting}

The \name\ pragma enables manual partitions to be used as well, as shown in Listing~\ref{code:partition}, which specifies two work regions.  The annotation is expanded to an internal function call that performs the partitioning and returns  a unique partition ID. In lines 6-7, the partition ID ({\tt part0}) is used to determine the work distributions of the two kernels.  This annotation allows more programmer control for optimal communication tuning and load balancing.


\subsection{\name\ Library Functions}
\label{ssub:api}
The library functions, with core APIs described in Table~\ref{tab:prototype}, encapsulate low-level details of the programming model.

\begin{table*}[t]
	\notsotiny
	\renewcommand{\arraystretch}{1.2}
	\centering
	\caption{Core \name\ interface library functions}
		\vspace{-8pt}
	\begin{tabularx}{\textwidth}{|X|X|}
		\hline
		\multicolumn{1}{|c|}{\textbf{Prototype}} & \multicolumn{1}{c|}{\textbf{Description}} \\
		\hline
		\funcproto{int \funcinit(int argc, char *argv[], char *kpath, char *dpath)} & 
		Initialize \name\ runtime environment and returns device ID.
		Take path to OpenCL kernel file ({\tt kpath}) and optional device information file ({\tt dpath}).\\
		\hline
		\funcproto{void \funcexit(void)} & 
		Terminate the \name\ runtime environment.\\
		\hline
		\funcproto{void \funcshow(int devID)} & 
		Display device information. Take device ID.\\ 
		\hline	
		\funcproto{\name\_t *\funccreate(char *sym, char *type, void *uA, int dim, ... )} & 
		Allocate host and device buffer and returns \name\ handle. 
		Take name, type, address, and size of user array ({\tt uA}).\\
		\hline
		\funcproto{int \funcpartition(\partitiontype\ type, int dim, ... )} & 
		Partition work item regions. Take type of partition and variable list for  array size and region for each dimension.  Supported types: {\tt ROW}, {\tt COL}, and {\tt BLOCK}.\\
		\hline
		\funcproto{void \funcapplykernel(char *kName, int partID,...)} & 
		Perform communication and kernel execution. 
		Take the kernel name, partition ID, and kernel arguments. \\
		\hline
		\funcproto{void \funcread(\name\_t *hA, void *uA,int partID)} \newline 
		$\Rightarrow$ Same for \funcwrite & 
		Read(Write) array section specified by partition ID from(to) \name({\tt hA}) to(from) user array ({\tt uA}).\\
		\hline
		\funcproto{void \funcreduce(\name\_t *hA, void *res, REDUCE\_OP op, int partID)} & 
		Reduce specific array sections of \name({\tt hA}) to a scalar value({\tt res}). Supported ops: {\tt SUM}, {\tt PROD}, {\tt MAX}, and {\tt MIN}.\\
		\hline
		\funcproto{void \funcabsuse(char *kName, int partID, \name\_t *hA, int devID, int dim, ... )} \newline 
		$\Rightarrow$ Same for \funcabsdef & 
		Set absolute section used(defined) for each device. The absolute section becomes LUSE(LDEF). \\
		\hline
		\funcproto{void \functrapezoiduse(char *kName, int partID, \name\_t *hA, int devID, int dim, ... )} \newline 
		$\Rightarrow$ Same for \functrapezoiddef & 
		Set predefined shape for LUSE(LDEF). Specify four positions of upper-left, upper-right, below-left, and below-right.\\ 
		\hline
	\end{tabularx}
	\label{tab:prototype}
	\vspace{-15pt}
\end{table*}

\paragraph{\funcinit} initializes the MPI and OpenCL systems and reads in data generated by the frontend.
A table is maintained for each \name 's use in a kernel call, which is initialized with information gathered by the frontend from  \emph{use} and \emph{def}.  This use/def and partitioning information are used to generate inter-process and host-device communication.
%
The last parameter of the \funcinit~function provides a device information file containing tuples \emph{(MPI rank, device ID)} to the kernel.  Note that MPI rankfile allows the MPI rank to be known before a program run.
The device ID can be used to specify which device(s) to use when multiple devices are available to a process. 

\paragraph{\funccreate} creates a host and device buffer for the \name\ on each process using \texttt{malloc()} and \texttt{clCreateBuffer()} respectively, and allocates and initializes the \name's GDEF.

\paragraph{\funcpartition} evenly partitions work item regions for each device for a given partitioning type. It then defines the partitioned work item region entry in a partition table 
maintained for the \name\ and return a unique partition ID that is used for work and data distribution.

\paragraph{\funcapplykernel} manages communication and launches a kernel.  It performs the following actions:
\begin{itemize}[leftmargin=*]
\item Bind arguments to the kernel call.  
The system finds the device buffer from \name\ handle, binds all other arguments directly, and calls the OpenCL function \texttt{clSetKernelArg()}.
\item Determine and perform communication.  {\em Use} \name s trigger communication. 
LUSE is updated by composing \emph{use} offset (in the kernel table) with partitioned work item regions, and the LUSE is intersected with GDEF to determine communication (Eqns.~\ref{eq:intersectionsend} and~\ref{eq:intersectionrecv}).  If the intersection resides in the device, it is transferred to the host using \texttt{clEnqueueReadBufferRect()}.
Then, any needed inter-process non-blocking (e.g., point-to-point or collective) communication is done, followed by transferring data to the device using \texttt{clEnqueueWriteBufferRect()}. 
Finally, the system updates the GDEF sets, as described in Eqns.~\ref{eq:update_gdef_send} and~\ref{eq:update_gdef_recv}.

\begin{sloppypar}
\item Execute the kernel. The preferred work-group size is found using \texttt{clGetKernelWorkGroupInfo()}, the work-group size is set, and the kernel is called using \texttt{clEnqueueNDRangeKernel()}.  Modified device buffers are known because of the {\em def} offset from the kernel table, and  
the system updates LDEF and GDEF (Eqns.~\ref{eq:update_gdef_send} and~\ref{eq:update_gdef_recv}) so that the host has the coherent copy.
\end{sloppypar}
\end{itemize}

\paragraph{\funcabsuse~and \funcabsdef} specify the absolute sections of the \name s annotated with \emph{use@} or \emph{def@} clauses (Section~\ref{subs:clauses}) which are then used to define LUSE and LDEF for each device.  LDEF updates must be precise because they affect GDEF, which defines who owns the coherent copy of a value. 
Multiple \funcabsdef~calls can be used to give precise updates, but for ease of programming and avoiding errors in entering each absolute section, \name\ supports predefined shapes for absolute section updates for LDEF (and LUSE).  For example, the \functrapezoiddef~function supports LDEF updates for two-dimensional trapezoidal or triangular shapes, which avoids multiple \funcabsdef~calls to update the LDEF. 

\paragraph{Utility library functions}
In addition to the core API functions, utility library functions are provided.  I/O utility functions allow the programmer to move data between user space arrays and \name s.  The \name\ runtime updates the GDEF, LDEF, and LUSE information to reflect the data movement, thus keeping these consistent with the actual memory state.  Reduction functions are also provided.  If all data for an \name\  is in the host memories, a local reduction followed by an MPI reduction is performed.  If some or all data is in the device memory, a device reduction is performed followed by an MPI reduction.

\subsection{A Case Study: Matrix Multiply}
\label{subs:usecase}

Listing~\ref{code:host} and~\ref{code:device} show a General Matrix Multiply (GEMM) implemented using \name. The program uses C host code and OpenCL device code to perform the matrix multiply $C = A\times B$ on three 1024$\times$1024 2D matrices.

\begin{lstlisting}[	caption={GEMM host code.},label={code:host}, frame=lines, breaklines=true, xleftmargin=2em, framexleftmargin=1.5em, numbers=left, numberstyle=\tiny, numbersep=5pt, captionpos=b, float=htb, aboveskip=0em, belowskip=-2em]
void main(int argc, char *argv[]) {
  int ni = 1024, nj = 1024, nk = 1024;
  float a[ni][nk], b[nk][nj], c[ni][nj], alpha, beta;
  ... // initialize variables
  
  HDArrayInit(argc, argv, "gemm.cl", NULL);
  int part0 = HDArrayPartition(ROW, 2, ni, nj, 0, 0, ni, nj)
  
  HDArray_t *hA = HDArrayCreate("a", "float", a, 2, ni, nk);
  HDArray_t *hB = HDArrayCreate("b", "float", b, 2, nk, nj);
  HDArray_t *hC = HDArrayCreate("c", "float", c, 2, ni, nj);
  HDArrayWrite(hA, a, part0);
  HDArrayWrite(hB, b, part0);
  HDArrayWrite(hC, c, part0);
  
  HDArrayApplyKernel("gemm", part0, hA, hB, hC, alpha, beta, ni, nj, nk);    
  HDArrayRead(hC, c, part0);
  HDArrayExit();
}
\end{lstlisting}
\begin{lstlisting}[caption={GEMM device code.},label={code:device},frame=lines, breaklines=true, xleftmargin=2em, framexleftmargin=2em, numbers=left, numberstyle=\tiny, numbersep=5pt,	literate={_}{\textsmallunderscore}1, captionpos=b, float=htb, belowskip=-3.5em] 
#pragma hdarray use(A,(0,*)) use(B,(*,0)) def(C,(0,0))
__kernel void gemm(__global float *A, __global float *B, __global float *C, 
                   float alph, float beta, int ni, int nj, int nk) {
  int i = get_global_id(1), j = get_global_id(0);
  if ((i < ni) && (j < nj)) {	
     C[i * nj + j] *= beta;
     for(int k=0; k < nk; k++)
        C[i*nj+j] += alph * A[i*nk+k] * B[k*nj+j];
  }
}
\end{lstlisting}
%

Line 6 of the host code initializes the MPI and OpenCL environments and finds available devices to run the OpenCL kernel implemented in ``{\tt gemm.cl}''. 
Line 7 evenly partitions the highest dimension of 2D array domain with regards to the number of devices. The function returns a partition ID, {\tt part0}, which represents the partitioned region and is used throughout the program. 

On lines 9-11, the host creates \name s and allocates host and device buffers with the same size of user-space arrays. 
After the allocation, the host binds program array variables ({\tt a}, {\tt b}, {\tt c}) to handles ({\tt hA}, {\tt hB}, {\tt hC}) that point into structures in the \name\ runtime and allow users to access device buffers holding data for their respective program arrays.
Lines 12-14 write user arrays into the device buffer of \name s according to the {\tt part0} specification. Therefore, the data is distributed to different devices.

On line 16, the host launches the ``gemm'' kernel using the {\tt part0} and kernel arguments. {\tt part0} is used for work distribution. As shown in Fig.~\ref{fig:applykernel}, the runtime then binds \name\ handles and host variables to the kernel arguments, handles necessary communication, and invokes the kernel (Listing~\ref{code:device}).
Line 17, reads the result of the computation from the device memory into user array {\tt c}. 
Finally, the host frees all the resources, including \name 's, and finalizes the parallel program in line 18. 

The device code in Listing~\ref{code:device} shows an ordinary OpenCL kernel to be called, with an annotation added on line 1. The annotation is a {\tt \#pragma \MakeLowercase{\name}} statement with offset clauses discussed in Section~\ref{subs:clauses}. These offsets, relative to a work item index, specify slices of the {\tt A}, {\tt B}, and {\tt C} arrays that are used and defined.
The code informs the runtime system that a single thread reads all elements of the row of the array {\tt A} and all elements of the column of the array {\tt B}. The zero offset indicates that each thread writes the result of the multiplication to its work item index of the array {\tt C}.

With the work partitioning ({\tt part0}) and per-thread array element access (offset) information provided by the host and device code, respectively, the runtime is able to generate LUSE and LDEF, and distribute work for the kernel.

%% file: implemention.tex
This section describes some implementation details not covered in Section~\ref{subs:ui}.  

 \subsection{Frontend and Execution Phase}
\label{subs:front_exec_phase}

 The frontend phase, shown in Fig.~\ref{sf:sysA}, uses a simple parser that performs three tasks: (1) parse OpenCL kernel functions and \name\ pragmas, (2) collect information, including {\em use} and {\em def} offset information, that is written to the file $ M $ (Fig.~\ref{fig:overview}) used to initialize the \name\ table, and (3) generate code for \name\ pragmas and directives that pass partitioning information to the runtime.
%
%

\begin{figure*}[htb]
	\centering
	\vspace{-15pt}
	\subfloat{\includegraphics[width=4in]{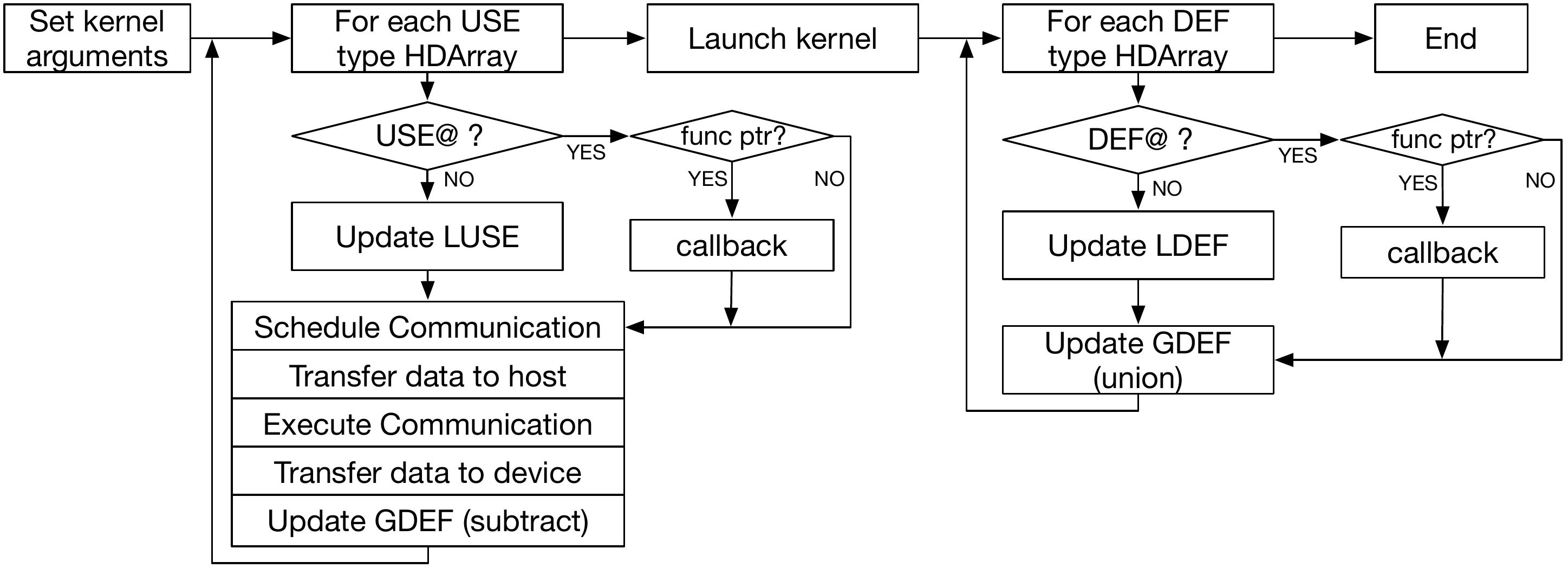}}%
	\vspace{-10pt}
	\caption{Logic of \funcapplykernel~function.}
	\label{fig:applykernel}
	\vspace{-30pt}
\end{figure*}

Fig.~\ref{sf:sysB} shows the execution phase. The main tasks of the \name\ execution phase are (1) maintaining information about \name\ sections residing on hosts and in devices, (2) determining and scheduling communication to ensure up-to-date data is available for computations, and (3) launching kernel executions.  
Fig.~\ref{fig:applykernel} sketches the logic of \funcapplykernel~which is the core part of the execution phase. 
%
%
%
%
%
%

\subsection{Reducing Runtime Overhead}
\label{subs:optimization}

The major overhead incurred by our baseline runtime system comes from intersecting and updating array sections to determine communication. Data to send or receive are found by intersecting GDEF with LUSE sets, as shown in Eqns.~\ref{eq:intersectionsend} and~\ref{eq:intersectionrecv}, which requires the number of computations on each process $ p $ that is linear in the number of processes.
To reduce the overhead, the \name\ system attempts to evaluate the intersections only when it is necessary to do so.  To assist in this, the system caches the last GDEF, LDEF, and LUSE sets as well as intersections per kernel call/partition ID. 
If LUSE and GDEF are unchanged for repeated kernel calls with the same partition ID, the system reuses the intersections from the last kernel call. 

%
%
%
LUSE and LDEF sets, accessed in a repetitive kernel call with the same partition ID, can be reused.  However, GDEF sets for that kernel call can change as they are a function of the LUSE and LDEF sets \emph{and} all previous GDEFs. Therefore, the system must check whether the GDEF sets have changed.  
An exact comparison of GDEFs takes O($n^2$) time, but the GDEF comparison overhead is reduced in two steps. First, the system maintains history buffers of the IDs of LDEF and LUSE sets for each \name\ that tracks LDEF/LUSE IDs for the entire program. It then evaluates the def-use chains in the buffer to determine changes in GDEF. For example, the system compares the last LDEF and LUSE IDs with the current LDEF and LUSE IDs. If the two pairs are the same, GDEF sets for last and current kernels are also the same because a GDEF update is a function of LDEF and LUSE, thereby the system bypasses the GDEF comparison. 
If the history buffer does not provide enough information, as the second step, the system performs an O($n$) comparison of GDEF sections, enabled by 
keeping the GDEF sections in sorted order when updated in Eqns.~\ref{eq:update_gdef_send} and~\ref{eq:update_gdef_recv}. The sorted GDEFs allow simple and linear-time GDEF comparisons.   

Finally, the system hides the overhead of section updates by overlapping the updates with host communication and device computation.  
It updates GDEF sets for \name~used during the communications which are all non-blocking. For a defined \name, the system updates LDEF and GDEF sets during the non-blocking kernel execution.

%% file: eval.tex
In this section, we evaluate the effectiveness of the proposed techniques with six publicly available benchmarks. 
%
Our evaluation is done using up to 32 OpenCL devices (limited by Xsede job submission policies) on the Xsede Comet cluster~\cite{towns2014xsede,Moore:2014:GDC:2616498.2616540}. Comet has 1,944 compute nodes and 72 GPU nodes, connected by a 56 Gbps FDR Infiniband. Each compute node consists of two 12 core Intel Xeon CPU E5-2680 processors running at 2.50 GHz, 128 GB of main memory. The GPU nodes consist of 36 NVIDIA P100 nodes and 36 NVIDIA K80 nodes, and each node has 4 GPUs. We use both P100 and K80 GPU nodes, each of which has total 40GB of device memory and utilize 4 GPU devices per node.

We use six micro-kernel benchmarks: GEMM, 2MM, 2D Convolution, Jacobi, Covariance, and Correlation from the Polyhedral Benchmark Suite for GPUs and accelerators (PolyBench/ACC)~\cite{6339595}.  We compile the benchmarks with gcc 4.9.2 with -O3 and use OpenMPI version 1.8.4. OpenCL version 1.2 is used to support NVIDIA devices.  Baseline numbers are found using the implementations provided by the benchmarks.  For \name\ system numbers, the OpenCL device code is augmented with \name\ pragmas with \emph{use} and \emph{def} clauses, and C host code that includes \name\ library calls and \emph{partition} clauses are added.
%

\begin{figure*}[t]
	\centering
	\subfloat{\includegraphics[width=4.5in]{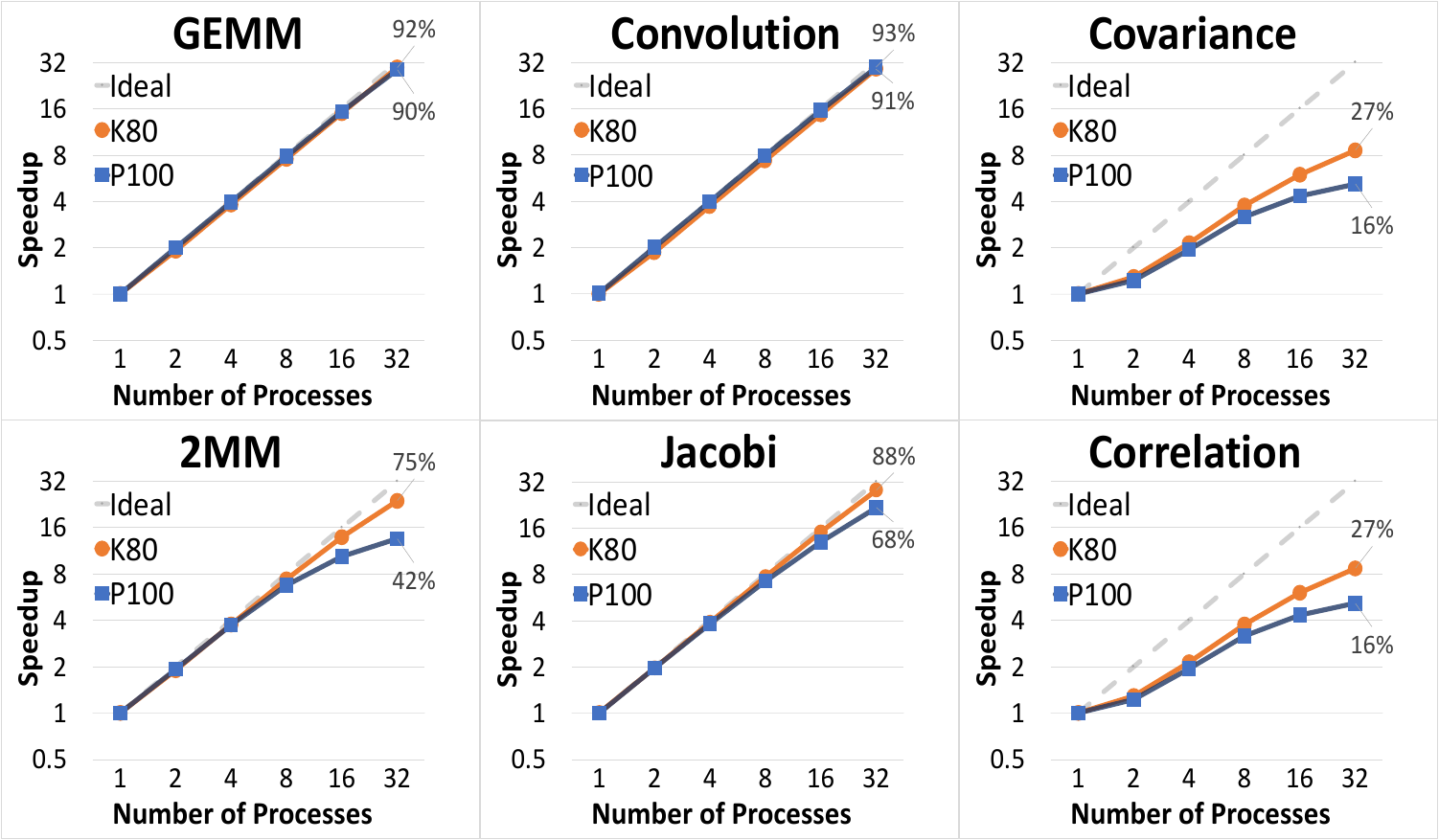}}%
	\vspace{-10pt}
	\caption{Scalability for the \name\ runtime system on P100 and K80 nodes. We show the speedup for each benchmark, which is the ratio of the execution time of a single device to the execution time of the number of devices indicated on the x-axis. All the benchmarks use an automatic row-wise partitioning for data and work distribution.}
	\label{fig:eval_scalability}
	\vspace{-15pt}
\end{figure*}

\subsection{Scalability}
\label{subs:eval_scalability}
%
%
Fig.~\ref{fig:eval_scalability} shows strong scaling on both P100 and K80 nodes. The baseline time is for running one OpenCL device without \name.  All the benchmarks perform a row-wise partition using the \funcpartition~function with a ROW argument for work and data distribution. Most benchmarks running on K80 nodes scale better than on P100 nodes because the P100 is faster than the K80, and thus the communication overhead on P100 nodes is a larger fraction of the computation time. 


GEMM, shown in Section~\ref{subs:usecase}, uses 10,240$\times$10,240 matrices with 100 iterations. The \name\ runtime system detects and generates all-gather collective communication because each OpenCL work-item needs row and column elements of arrays for computation. Scaling is good to 32 processes, with similar efficiencies on the K80 (92\%) and P100 (90\%), due to the low ratio of communication to kernel execution time.
2MM performs two matrix multiplications, $D = A \times B$ followed by $E = C \times D$. It differs from GEMM in that 2MM runs two kernel functions within a loop and exhibits a data dependency because one kernel defines the array $D$ used by the other kernel. With the row-wise partitioning, the efficiency drops off to about 75\% (42\%) on the K80 (P100) at 32 processes because of the communication cost. The cost is proportional to the number of processes, and every iteration requires the communication: once for the array $B$, and 100 times for the array $D$.   

\begin{figure*}[t]
	\centering
	\subfloat{\includegraphics[width=3.2in]{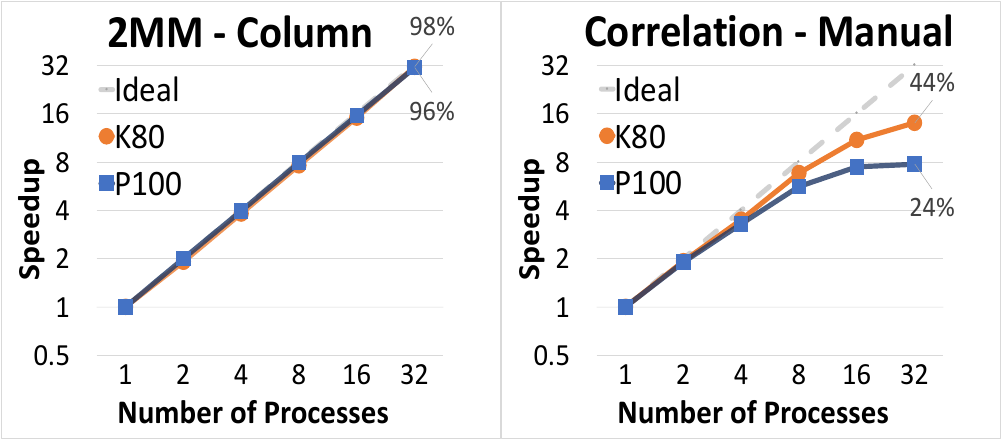}}%
	\vspace{-10pt}
	\caption{Scalability for the \name\ runtime system with different partitioning methods. 2MM uses automatic column-wise partitioning and Correlation uses manual row- and column-wise partitionings.}
	\label{fig:scalability_diff_partition}
\end{figure*}

A different partitioning can be used to reduce the communication cost. 2MM with column-wise partitioning, as shown in Fig.~\ref{fig:scalability_diff_partition}, only communicates twice for arrays $A$ and $C$, and the efficiency is about 98\% (96\%) on the K80 (P100) at 32 processes. Table~\ref{tab:partitioning} shows communication volumes of all 32 processes and noticeable volume difference for 2MM.  This performance tuning was done by simply changing the \partitiontype\ argument of \funcpartition~function.

\begin{table}[t]
	\scriptsize
	\renewcommand{\arraystretch}{1.3}
	\centering
	\caption{Total communication volume for 32 processes}
	\label{tab:partitioning}
	\vspace{-8pt}
	\begin{tabular}{|c|c|c|c|c|c|c|}
		\hline
		Partition & Convolution   & JACOBI       & GEMM     & 2MM          & Covariance & Correlation \\
		\hline
		Default (Row)        & 5 MB                & 473 GB        & 12  GB     & 1262 GB    &  1268 GB     &  1268 GB      \\
		Customized  & 5 MB                & 473 GB         & 12 GB     & 25 GB        &  811 GB     &   811 GB      \\
		\hline
	\end{tabular}
	\vspace{-15pt}
\end{table}

Both Jacobi and Convolution kernels are iterative stencil codes, with four and eight neighbors, respectively.  The offsets of \emph{use} and \emph{def} clauses have similar patterns for both benchmarks. For Jacobi, the device code consists of two kernels. One kernel processes the following computation: 
{
\small
\begin{align*}
A[i][j]=(B[i][j-1]+B[i][j+1]+B[i-1][j]+B[i+1][j]) / 4
\end{align*}
}%
The \emph{use} clauses are specified for the kernel with four offsets, (0,-1), (0,+1), (-1,0), (+1,0), for an array $B$. The other kernel performs $B[i][j]=A[i][j]$, and zero offsets are used. The host code allocates user space arrays to have ghost cells at the array boundary, and then generates two partition IDs: one that partitions the entire region for data distribution and the other that excludes the cells for work distribution. Two kernels have a data dependency on array $B$ in the iteration space. Convolution has four additional offsets added, but there is no data dependency. Both kernels use 20,480$\times$24,080 matrices with 100,000 iterations, and the runtime detects and schedules a point-to-point communication. 
Both benchmarks scale well with an efficiency of 88\% (68\%) on the K80 (P100) for Jacobi, and 91\% (93\%) on the K80 (P100) for Convolution at 32 processes. Similar to GEMM, two nodes for Convolution show similar efficiencies due to the small communication overhead.
%
%

Covariance and Correlation are data mining benchmarks that compute a measure from statistics that show how linearly related two variables are. These benchmarks have triangular-shape array accesses, requiring the absolute section interface discussed in Section~\ref{subs:absclause}.
Both use 10,240 vectors and 10,240$\times$10,240 matrices with 100 iterations, and the system detects point-to-point and all-gather communication. 
Scaling is poor with the default row-wise partitioning with an efficiency of 27\% (16\%) on the K80 (P100) for Correlation (similar to Covariance).
This is because evenly distributing work using \funcpartition~causes poor work and communication load balancing for kernels that have triangular access patterns. The most time-consuming computation is done from the upper-triangular section of an array which later requires communication to make the array symmetric. As a result, each device gets a different amount of work, and a device with the most computation also has the most communication, which leads to the imbalance of computation and communication across the devices.

Manual partitioning (Listing~\ref{code:partition}) with optimized absolute section updates to balance the work and communication among devices, gives better scalability with an efficiency of 44\% (24\%) on the K80 (P100) with the reduced communication volume as shown in Fig.~\ref{fig:scalability_diff_partition} and Table~\ref{tab:partitioning}, respectively. This result highlights the value of integrating manual and automatic partitioning.
Also, the performance tuning does not require any changes in kernel code, but only a few lines are changed in absolute section updates and partitioning in  the host code.

\subsection{Runtime Overhead}
\label{subs:eval_overhead}

\begin{figure}[tb]
	\centering
	\includegraphics[width=4in]{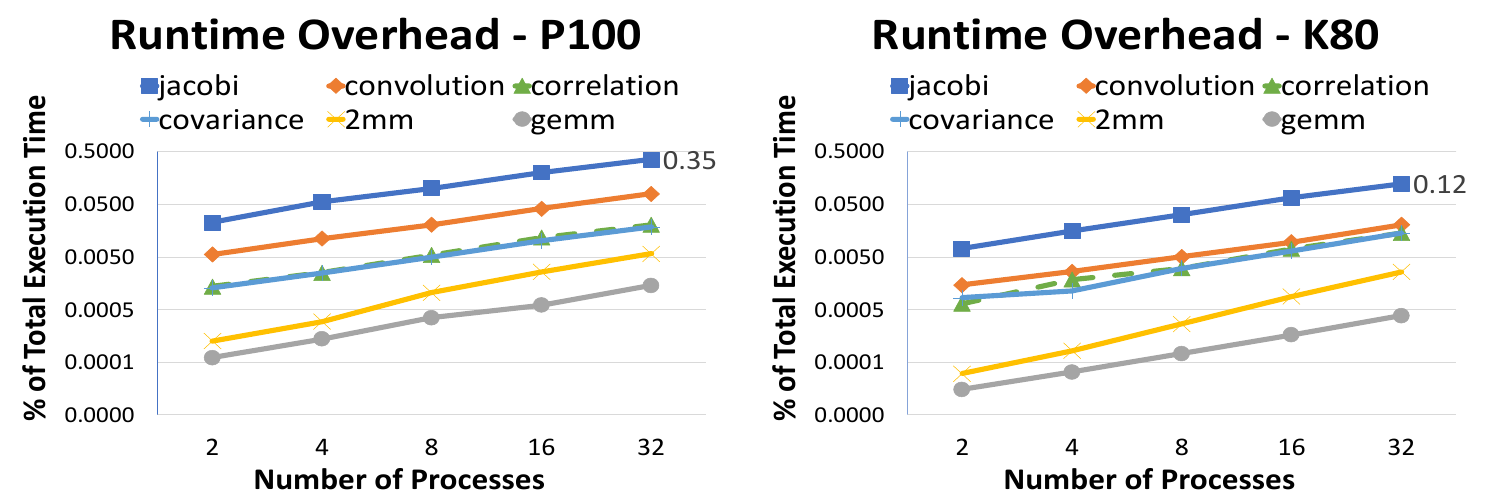}
	\vspace{-10pt}
	\caption{Total runtime overhead for all six benchmarks on both P100 and K80. The highest overhead is 0.35\% from Jacobi on P100.}
	\label{fig:overheadall}
	\vspace{-10pt}
\end{figure}
\begin{figure}[tb]
	\centering
	\includegraphics[width=4in]{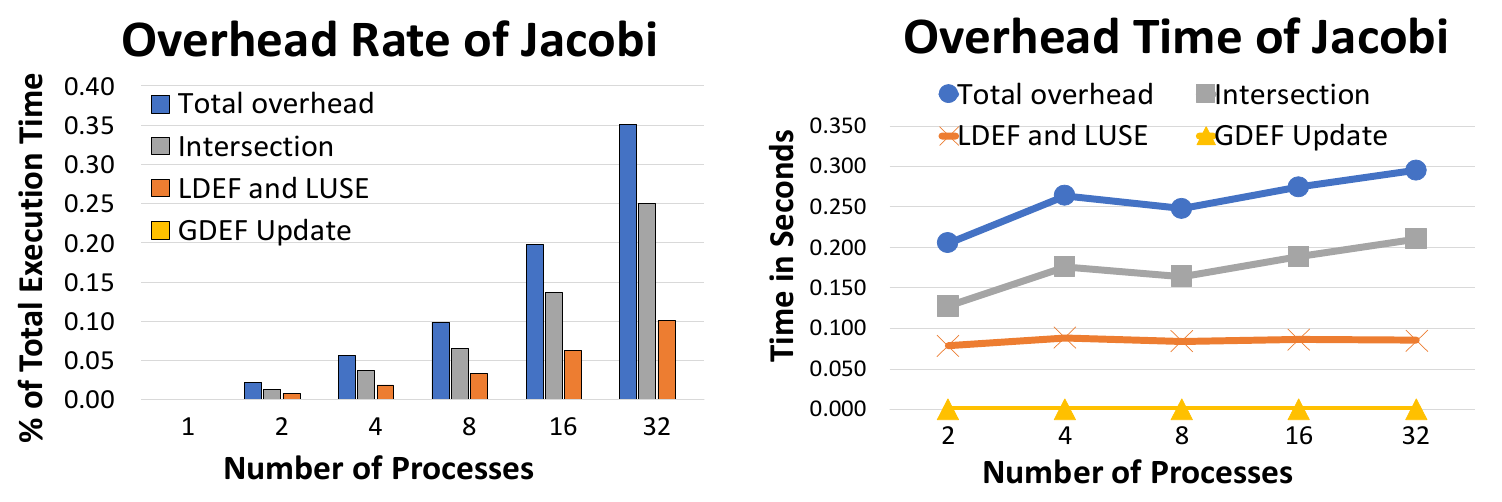}
	\vspace{-10pt}
	\caption{Breakdown of runtime overhead for Jacobi on P100. There is no overhead of GDEF updates  because these updates are overlapped with communication and computation. Running a single device also does not incur any overhead.}
	\label{fig:jacobi_overhead}
	\vspace{-13pt}
\end{figure}

Fig.~\ref{fig:overheadall} shows the percentage of runtime overhead based on total execution time. All of the benchmarks have less than 0.36\% overhead on up to 32 processes. The benchmarks on a P100 node have more overhead than on a K80 as the P100 is faster than the K80, similar to the scalability difference. Our effort to minimize the runtime overhead, as discussed in Section~\ref{subs:optimization}, significantly reduced the cost of calculating GDEF, LDEF, LUSE, and intersection for all the benchmarks. Without the optimization, the \name\ baseline system suffers from the overhead of section calculations, which increases proportional to the number of processes.

Breaking down the highest overhead benchmark in Fig.~\ref{fig:jacobi_overhead}, the overhead of GDEF updates are zero because they overlap with, and finish before, the communication and kernel computation. This is a large improvement because every kernel call requires the GDEF update.
The intersection overhead, which includes both intersection time and caching time, is much smaller, e.g., by a factor of 19 for Jacobi on P100, than the baseline system. The result shows the benefit of using the LDEF/LUSE history buffer and linear-time GDEF comparison. 
Caching LDEF and LUSE for each kernel call is also beneficial. The LDEF and LUSE update overhead does not linearly increase in time because the local sections are reused after the first iteration of the kernel call. 
Finally, although not shown in the figure, the system reduces the number of sections by merging adjacent or redundant sections, further reducing the overheads of intersecting.

%% file: related.tex
Our paper is related to previous efforts to simplify distributed accelerator programming with runtime support for efficient communication.
%
\label{subs:related_acc_cluster}
%
Hydra~\cite{sakdhnagool2015hydra} is a compiler-assisted runtime system which extends OMPD~\cite{Kwon:2012:HAO:2145816.2145827}'s hybrid compiler-runtime communication analysis to  translate and execute OpenMP programs on accelerator clusters. One difference is that \name\ handles communication without any static analysis, allowing programmers to use separate compilation and external binary libraries.

PGAS languages~\cite{upc2005upc, numrich1998co, charles2005x10}
have been extended to support accelerator clusters~\cite{Lee:2011:EXP:2238356.2238410, potluri2013extending, 7081675}.
PGAS languages expose a global shared array to relieve programmers from data distribution and communication handling but require the  specification of the affinity between data and threads, which makes data owned by a thread. As the data ownership is strongly coupled with computation, changing data distribution may require the modification of computation code. Our approach gives more freedom to the programmer to re-distribute data at any parallel program point without changes in kernel code. High Performance Fortran (HPF)~\cite{RiceUniversity:1993:HPF:174223.158909, Gupta:1995:HCI:224170.224422}
does not support accelerators.

Researchers have proposed language extensions for the programmability of heterogeneous clusters. SnuCL~\cite{kim2012snucl} and SnuCL-D~\cite{kim2016distributed} enable OpenCL applications to run in a distributed manner without any modification. 
dCUDA~\cite{gysi2016dcuda} automatically overlaps on-node computation and inter-node communication with hardware support and device-side remote memory access operations. It combines the MPI and CUDA programming models into a single CUDA kernel. IMPACC~\cite{kim2016impacc} integrates MPI and OpenACC~\cite{openacc} while exploiting shared memory parallelism. It reduces the communication cost through unified MPI communication routines, a unified node virtual address space, node heap aliasing technique, etc. Despite their optimized communication with little or no code changes, programmers are forced to manage numerous low-level details of the accelerator or MPI programming because these tools provide an abstraction level analogous to OpenCL/CUDA or MPI, and require explicit data transfer or communication code.

OmpSs~\cite{bueno2012productive} supports task parallelism and directives for computation offloading and communication handling. Programmers specify accessed regions of shared data, but no convenient way to define and operate on subarrays is provided. We differ by supporting data parallelism and allowing users to specify per-thread offset information, 
and both work and data partitioning can be done automatically or manually.
HOMP~\cite{yan2017homp} proposes an extension of OpenMP for distributing and binding computation and data, which gives users more control of managing data and computation, but 
lacks cluster support and manual partitioning for specific devices.

Vi{\~n}as \textit{et al}.~\cite{vinas2016towards} proposed the hybrid use of Hierarchically Tiled Array (HTA)~\cite{bikshandi2006programming} for globally distributed arrays and Heterogeneous Programming Library (HPL)~\cite{vinas2013exploiting} for accelerators. Both HTA and HPL C++ libraries provide implicit parallelism and communication and hide many low-level details of MPI and OpenCL; however, there exist two different arrays: an HTA and an HPL Array, which programmers need to define and maintain. Explicit data transfer from the HPL Array to an HTA is also necessary.
PARRAY~\cite{Chen:2012:PUA:2145816.2145838, cui2015programming} is a C language extension that introduces novel array types to separate the logical and physical structure and what kind of process/thread will operate on a dimension. 
Unlike \name, users need to specify communication mechanisms for every array and explicitly insert communication code. 

Skeleton libraries~\cite{ernsting2017data, majeed2013cluster} 
differ from us in that they can only support applications in which all the computational patterns are covered by the skeletons.
%
Other popular platforms such as
NumbaPro~\cite{numbapro}, Arrayfire~\cite{arrayfire}, PyCUDA~\cite{KLOCKNER2012157},  Copperhead~\cite{Catanzaro:2011:CCE:1941553.1941562},  
OpenACC
and 
OpenMP 4.0 %
all aim to make accelerator programming easier, but only target a single node.

%% file: conclusion.tex


We have presented the \name\ interface and runtime system for accelerator clusters. The interface features a novel global programming model which separates work partitioning from the concept of data distribution, thus enabling straightforward and flexible work distribution.

The interface abstracts away many low-level details of multiple address space programming, yet supports a low-level array programming environment through the \name\ annotations and APIs for performance tuning. We showed how the \name\ interface could help programmers to write and tune array-based programs for distributed devices. The offsets provide an intuitive and simple way to describe the access patterns of kernels, and the patterns can be easily changed by simply adjusting partitions without the modification of kernel code.

The \name\ runtime system performs efficient and fully automatic communication by managing the array sections. We presented optimizations including the caching mechanism and communication and computation overlap, which reduce or hide much of the overheads of communication detection.

Future work being considered is the ability to adjust work partitions assigned to devices.  This capability would allow splitting up computations to fit them in small device memories, and adjusting the sizes of work assigned to heterogeneous devices to provide load balancing.



\if{0}
\subsection{Future Work}
\label{subs:design_discuss}
With additional enhancements, the \name\ can be more efficient and reliable with a wider range of array-based programs. We discuss these now.

\subsubsection{Handling Applications that Exceed Device Memory}
Unlike CPUs, accelerators typically have small physical memories and no virtual addressing.
The \name\ program will fail if the available device memory size is smaller than needed for computation because it allocates the entire array when \name s are created and uses the entire partitioned work item region for a kernel. To fit in the device memory, we can allocate the device buffer during the kernel computation and use a host buffer as a backing store to split the partitioned region to work with smaller data. This optimization is feasible by enhancing \funcapplykernel.

\subsubsection{Automatic Utilizations of Computational Resources}
The \name\ runtime does not support automatic load balancing. 
For \emph{repetitive} programs, in which the communication pattern is the same in all iterations of the serial loop enclosing kernel calls, the runtime can balance the work by analyzing the performance of devices at each iteration and dynamically adjust work and data distributions. Another utilization such as overlapping computation and communication is not also supported because of the SPMD execution model we use. The design of providing the overlap is ongoing future work.

\fi